# Towards an Understanding of Scaling Frameworks and Business Agility

## A Summary of the 6th International Workshop at XP2018

Torgeir Dingsøyr
SINTEF
Norway
torgeird@sintef.no

Nils Brede Moe
SINTEF
Norway
nils.b.moe@sintef.no

Helena Holmström Ohlsson
Malmö University
Sweden
helena.holmstrom.olsson
@mah.se

## ABSTRACT

Large development projects and programs are conducted using agile development methods, with an increasing body of advice from practitioners and from research. This sixth workshop showed in increasing interest in scaling frameworks and in topics related to achieving business agility. This article summarizes four contributed papers, discussions in "open space" format and also presents a revised research agenda for large-scale agile development.

## 1 INTRODUCTION

This is a brief summary of the 6th International Workshop on Large-Scale Agile Development which was conducted as a half-day workshop at the International Conference on Agile Software Development, XP2018. The workshop received six submissions, of which four were selected for presentation at the workshop. We briefly summarize articles presented, discussions in "open space" format, and finally a revision of the research agenda for large-scale agile development. Summaries from previous workshops can be found here [1-4].

## 2 SUMMARY OF CONTRIBUTIONS

The articles accepted for presentation at the workshop focused on selecting a scaling framework, coordination in large-scale development, portfolio management and finally architectural models:

### 2.1 Selecting a Scaling Framework

Diebold, Schmitt and Theobald [5] recognizes that even though there are a number of existing frameworks for how to scale agile development practices, only a few of these are used and there is a lack of guidelines that help practitioners select the most appropriate one. Based on a literature review of existing frameworks, as well as workshops with practitioners, the authors present a set of criteria that can be used to compare different frameworks and that help selecting the most suitable one for a specific context. In their research, the authors identify four categories consisting of a total of 25 different criteria for comparison. Examples of these criteria are program structure, degree of flexibility, technical practices, certification and coaching efforts. Based on a selected set of these criteria, the authors compare 12 different agile frameworks and concludes that companies benefit from first using the high-level categories in order to identify a selected set of suitable frameworks. Next, and for the selected frameworks, a more detailed comparison can be done based on the underlying scaling and agile practices and the more elaborate criteria.

### 2.2 Coordination in Large-Scale Agile

Stray [6] described how agile projects, teams and team members coordinate by using feedback such as ad-hoc conversations, scheduled meetings and unscheduled meetings. Based on a survey of 65 members working in large-scale agile projects in Poland, Norway and China, the author concludes that on average, an agile team typically consists of eight members, the participants spend on average 1.1 hours per day in scheduled meetings, 1.6 hours in unscheduled meetings and ad-hoc conversations and a total of 13.5 hours per week in coordination by feedback. Furthermore, the majority of the respondents had team members distributed across sites and the survey results show that the team size of distributed teams is significantly larger than that of co-located teams. In addition, the paper proposes a theory for coordination by feedback and suggests future research for validating the theory by assessing scheduled and unscheduled meetings in relation to the different types of value they provide.

### 2.3 Agile Portfolio Management

Horlach, Schirmer, Böhmann and Drews [7] focused on the increasing challenge for companies to select a fitting approach for IT portfolio management in an organization consisting of agile teams with high autonomy. As recognized by the authors, it is unclear what 'agile' means for portfolio management and what principles are underlying portfolio management in large scale agile organizations. Based on design science, the authors present patterns for agile portfolio management, i.e. basic principles such as processes, methods, roles and artifacts for portfolio management. With the portfolio management patterns, the authors seek to enable agile teams to self-organize their work and service delivery while at the same time being able to conform to e.g. regulatory issues.

### 2.4 Architectural Models

Santos, Pereira, Morais, Barros, Ferreira and Machado [8] recognizes the lack of requirements and architectural modeling support in many of the agile development practices. Typically, requirements management and architectural modeling practices are considered as a heavy documentation processes and therefore, they are addressed lightly but not sufficiently. Based on research aiming at platform development, the authors describe an approach capable of deriving logical architectures in order to establish the initial requirements that can be passed on to agile development teams. The approach proposes a systematic transformation of model-based requirements, i.e. UML Use Cases and Component diagrams, into ASD-oriented requirements according to backlog items such as e.g. themes, epics and user stories. Complexity is addressed by scaling the development to include distributed Scrum teams and the proposed architecture is used to modularize, refine, and use as input for the backlog structure.

## 3   OPEN SPACE DISCUSSIONS

The following main topics were discussed during the workshop: Design thinking in large-scale agile, meetings in large-scale agile, enterprise architecture and implementing SAFe.

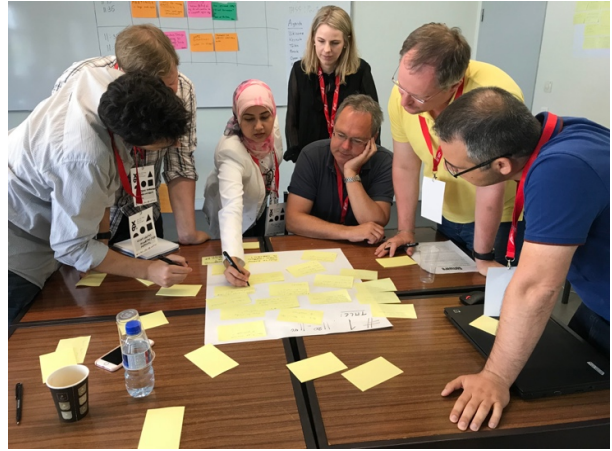

**Figure 1: Workshop participants in one of the open space discussions.**

### 3.1 Design thinking and Large-Scale Agile

Design thinking is about exploring and solving problems in a structured way. Design thinking refers to creative strategies used during the process of designing a system, e.g. when defining the product roadmap or as a process for identifying topics for a Hackathon. In a large-scale setting when many teams are involved in the development work, several challenges emerge regarding how to organize the creative work of designing the system. Examples are: when to do the design work, how to organize the design work and who to involve. Should the whole development team be involved, part of the team or is it only the designer/design team with architects that should do the creative work? If the development teams are not involved there might be a handover problem between the designers, architects and the team. However, if the development teams use a lot of time in the creative design process they will have less time to do the implementation work. Research questions might be:

How to involve the development team, designers and architects in design work in a large-scale setting?

Which problems are suited for a design thinking approach (e.g. only known unknown problems)?

When and how frequent should you do design work in a large-scale setting?

### 3.2 Meetings in large Scale-agile

Agile software development relies primarily on coordination by feedback (mutual adjustment). Scheduled and unscheduled meetings are important for coordination by feedback in large scale agile projects and programs. While meetings are essential for coordination, several challenges emerge in a large-scale setting: Too many meetings reduce the time for doing the actual work, there is often a confusion regarding who should participate in which meetings, people are missing good meeting criteria, meetings can be used for micromanagement when leaders ask people to give detailed status reporting. Research questions might be:

> How to moderate good agile meetings in a large-scale setting?
> What are good meeting criteria?
> How to study meetings?
> How to reduce the number of meetings in large-scale agile.
> How can technology support meetings in large-scale agile?

### 3.3 Enterprise Architecture

Every software system has an architecture because every system can be shown to be composed of elements and relations among them. The architecture of a large-scale system embodies information about how the elements relate to each other. In large-scale agile one main challenge is to balance the need for an upfront vs. emergent architecture. If the architecture is not planned – there might be chaos, however if it is too much planned the project might end up as a waterfall project. Further, understanding how to involve the enterprise architect is a challenge, because the role of the enterprise architect differs based on the maturity of the product, the teams and organization. Research questions might be:

- When and how to involve the enterprise architect in the work and in the teams
- How much upfront modelling is required?
- What decision power should the architect have vs the teams?

### 3.4 Implementing SAFe

Several frameworks for scaling agile has been created, such as the Large-Scale Scrum (LeSS) and Scaled Agile Framework (SAFe). There is a high attention on how companies can adopt and study SAFe. SAFe is perceived as a structured way of organizing the work, this includes, e.g., release trains with joint program increment planning days. Further, SAFe seems to create a structure with more organizational control, which might leave less flexibility for meetings to emerge and for teams to take the initiative for coordination. The SAFe framework is seen as a complicated framework to understand and a top-down approach to coordination in large-scale agile. While some countries have a high adoption rate of SAFe, some countries have a lower adoption rate. Some expressed that to succeed with SAFE there is a need to explain why you do it and to work on change management (inform and involve). The following questions were raised:

- What makes SAFE not agile?
- How large does a company need to be to get benefit from SAFe?
- What do you need to implement in SAFe to get the described benefits, and what is not that important?
- Why do some companies fail in adopting SAFe while others succeed?
- What contexts are better suited for implementing SAFe?
- What are the advantage of piking own practices vs adopting a full framework?

## 4 RESEARCH AGENDA

The 25 participants at the workshop were divided into tables which read the research agenda from 2017 [1] and were first asked to identify 2-3 new topics that were not on the agenda. The new topics were presented to all, and the groups were told to discuss priority and identify the "three most important topics" for research. The votes were collected from the groups, which gave a list of topics with "first", "second" and "third" priority:

Fist priority
- Agile in public/ IT government
- Agile transformation
- Business agility
- Scaling agile

Second priority
- Integrating non-software and software parts of the organization into agile (enterprise agile)
- Knowledge sharing / networks
- Patterns in large scale agile development (identify typical problems and provide appropriate patterns to address them)
- The role of architects and architecture in agile

Third priority:
- How DevOps affects agile
- Inter-team coordination

## 5 CONCLUSION

The sixth international workshop on large-scale agile development show a number of important topics for development projects and programs that are of critical importance to society and companies. As last year, the scaling frameworks are receiving interest and the topics with high priority in the research agenda indicates an increased interest in topics related to achieving business agility, such as "agile transformations" and "enterprise agility".

### Acknowledgments

We would like to thank members of the program committee: Steve Adolph (Development Knowledge, Canada), Julian Bass, (University of Salford, UK), Andrew Begel (Microsoft Research, USA), Finn Olav Bjørnson (Norwegian University of Science and Technology), Kevin Crowston (Syracuse University, USA), Siva Dorairaj (Software Education, New Zealand), Jutta Eckstein (IT communication, Germany), Davide Falessi (California Polytechnic State University, USA), Tomas Gustavsson (Karlstad University, Sweden), Elke Hochmüller (Carinthia University of Applied Sciences, Austria), Eric Knauss (Gothenburg University, Sweden), Maarit Laanti (Nitor Delta, Finland), Parastoo Mohaghegi (Norwegian Welfare Administration), Knut Rolland, (SINTEF Digital, Norway), Kai Spohrer (University of Mannheim,

Germany), Christoph Stettina (Leiden University, the Netherlands).

The work with the workshop was partially supported by the project Agile 2.0, supported by the Research council of Norway through grant 236759 and by the companies Equinor, Kantega, Kongsberg Defence & Aerospace, Sopra Steria, and Sticos.